\documentclass[a4paper,11pt]{article}
\pdfoutput = 1
\usepackage{jheppub}
\usepackage{amsmath,amssymb,amscd,braket,amsfonts}
\usepackage{color}
\usepackage{graphicx}
\usepackage{slashed}
\usepackage{amsthm}
\usepackage{verbatim}
\usepackage{relsize}

\newcommand{\bea}{\begin{eqnarray}}
\newcommand{\eea}{\end{eqnarray}}
\newcommand{\be}{\begin{equation}}
\newcommand{\ee}{\end{equation}}
\newcommand{\ba}{\begin{align}}
\newcommand{\ea}{\end{align}}

\title{
Thermal three-point functions from holographic Schwinger-Keldysh contours
}
\date{2022}

\author[a]{Christiana Pantelidou}
\affiliation[a]{School of Mathematics and Statistics, University College Dublin, Belfield, Dublin 4, Ireland}
\emailAdd{christiana.pantelidou@ucd.ie}

\author[b]{and Benjamin Withers}
\affiliation[b]{Mathematical Sciences and STAG Research Centre, University of Southampton, Highfield, Southampton SO17 1BJ, UK}
\emailAdd{b.s.withers@soton.ac.uk}

\abstract{
We compute fully retarded scalar three-point functions of holographic CFTs at finite temperature using real-time holography.
They describe the nonlinear response of a holographic medium under scalar forcing, and display single and higher-order poles associated to resonant QNM excitations. 
This involves computing the bulk-to-bulk propagator on a piecewise mixed-signature spacetime, the dual of the Schwinger-Keldysh contour.
We show this construction is equivalent to imposing ingoing boundary conditions on a single copy of a black hole spacetime, similar to the case of the two-point function.
We also compute retarded scalar correlators with stress-tensor insertions in general CFTs by solving Ward identities on the Schwinger-Keldysh contour.
}

\begin{document}
\maketitle

\section{Introduction}
Holography provides a non-perturbative real-time formalism for strongly coupled quantum field theories (QFTs). At finite temperature, the dynamics of the QFT are governed by those of an asymptotically-AdS black hole spacetime. In linear response these dynamics are universal, described by a spectrum of black hole quasinormal modes (QNMs). There is thus an intimate connection between QNMs and finite temperature QFT observables. This is made manifest in \cite{Birmingham:2001pj,Son:2002sd} in which QNM frequencies appear as poles of retarded two-point functions in the QFT.

It is natural to ask if this universality continues to nonlinear response properties of the holographic system. 
To investigate this we study higher-point correlation functions of the QFT. In particular we compute the so-called `fully retarded' correlation functions, which are expectation values of the R-product of operators in the thermal state, defined as\footnote{In the `r/a' notation this is precisely $G_{raa\ldots a}(x,x_1,x_2,\ldots, x_{n-1})$ at $n$ points as shown in \cite{CHOU19851} and as we review below. For a discussion of the R product and related operator orderings see \cite{Meltzer:2021bmb}. }
\bea
R[\mathcal{O}(x);\mathcal{O}(x_1) \ldots \mathcal{O}(x_{n-1})] \equiv && (-i)^{n-1}  \sum_{i} \theta(x^0-x_{i_1}^0)\theta(x^0_{i_1}-x_{i_2}^0)\ldots\theta(x^0_{i_{n-2}}-x_{i_{n-1}}^0) \nonumber\\
&&\qquad\left[\ldots\left[\left[\mathcal{O}(x),\mathcal{O}(x_{i_1})\right],\mathcal{O}(x_{i_2})\right]\ldots , \mathcal{O}(x_{i_{n-1}})\right],\label{Rproduct}
\eea
where the sum is over all permutations of the $x_i$ insertions.
In particular $\left<R[\mathcal{O}(x);\mathcal{O}(x_1)]\right> = -i \theta(x^0-x_1^0)\left<\left[\mathcal{O}(x),\mathcal{O}(x_1)\right]\right>$ is the familiar retarded two-point function, and \eqref{Rproduct} gives the natural extension of this object to higher points, corresponding to the causal response of the system at $x$ due to the insertions at $x_1,\ldots, x_{n-1}$. We focus primarily on scalar operators $\mathcal{O}$.

To compute these observables in holography it is necessary to go beyond the linear-response prescription of \cite{Son:2002sd, Herzog:2002pc} and turn to a fully real-time holographic formalism. We use the formalism of \cite{Skenderis:2008dh, Skenderis:2008dg, vanRees:2009rw} in which one constructs the bulk by `filling in' the real-time field theory contour of interest, resulting in a piecewise mixed-signature spacetime.\footnote{Another -- not necessarily distinct -- approach is based on restricting to ingoing coordinates and utilising the log branch point for the linearised scalar at the horizon to produce the two Lorentzian segments of the SK contour \cite{Glorioso:2018mmw}.} Thermal expectation values of \eqref{Rproduct} are naturally computed using a Schwinger-Keldysh (SK) contour in field theory. With the appropriate scalar insertions for the $n$-point function of interest, this contour fills in to become pieces of the Schwarzschild spacetime together with perturbative corrections capturing backreaction and interaction of bulk scalar fields. The perturbative corrections can be computed with the aid of the scalar bulk-to-bulk propagator defined on the piecewise Schwarzschild spacetime, which we construct in this work. This calculation shows that the expectation values of \eqref{Rproduct} can instead be computed by replacing the piecewise spacetime by a single Lorentzian copy of the black hole with ingoing boundary conditions at the horizon. This extends the analogous two-point computation in \cite{vanRees:2009rw} and confirms arguments presented there for higher points.

With this result established, we use it to explicitly compute the retarded scalar three-point functions numerically, in momentum space. The response of the system under driving probes bulk interactions and hence interactions of QNMs.\footnote{For bulk aspects of QNM interactions in AdS see \cite{Jansen:2020ign, Sberna:2021eui}.} For example, if the system is driven such that the sum of all driving momenta coincides with the momentum of a QNM mode, then the system exhibits a resonance. Correspondingly we find the three-point function contains a simple pole at this momentum. Other kinematical arrangements can be found such that more than one QNM is excited in the bulk leading to higher-order poles in the correlators. Two QNMs excited gives a order-two pole in the correlator, and so on. All of the singularities in the correlation functions we compute can be understood in precisely this way, thus their analytic structure is inherited from the universality of black hole ringdown, just like the two-point function. Such interactions of QNMs are likely to be of observational relevance in the asymptotically flat context \cite{Ioka:2007ak, Sberna:2021eui}.

\sloppy The final aspect addressed in this work is the effect of the system heating up due to external time-dependent driving by scalar fields. This process is captured by $\left<R[T_{tt}(x); \mathcal{O}(x_1)\mathcal{O}(x_2)]\right>$ which we compute from the diffeomorphism Ward identity on the SK contour. The result is expressed analytically in terms of $\left<R[\mathcal{O}(x); \mathcal{O}(x_1)]\right>$.

Related work computing thermal three-point functions can be found for CFT$_2$ in momentum space in \cite{Becker:2014jla}, and at large operator dimension in position space in \cite{Rodriguez-Gomez:2021mkk}. Three-point functions are also considered in \cite{Jana:2020vyx} together with a discussion of Witten diagrams on the piecewise bulk geometry. Scalar three-point functions with a $\lambda \phi^3$-interaction have also been constructed in vacuum using the bulk dual of the in-out real-time prescription \cite{Botta-Cantcheff:2017qir}. Real-time holography has also been used to investigate bulk excited states, where insertions or coherent sources are introduced in the Euclidean segment \cite{Botta-Cantcheff:2015sav, Christodoulou:2016nej, Marolf:2017kvq, Botta-Cantcheff:2018brv, Botta-Cantcheff:2019apr, Chen:2019ror, Arias:2020qpg, Belin:2020zjb, Martinez:2021uqo}. In addition, real-time holography has been employed extensively in relation to hydrodynamic or derivative-expanded effective theories \cite{deBoer:2018qqm, Glorioso:2018mmw, Jana:2020vyx, Loganayagam:2020eue, Loganayagam:2020iol, Ghosh:2020lel, He:2021jna, He:2022jnc, He:2022deg}. Finally, this approach was also employed in the context of heavy quarks moving in a strongly-coupled plasma in \cite{Chakrabarty:2019aeu}, where non-linear corrections to the Langevin effective action were computed.
\\\\
The paper is organised as follows. In section \ref{sec:fieldtheory} we review how thermal expectation values of \eqref{Rproduct} are obtained from a field theory path integral on the Schwinger-Keldysh contour. In section \ref{sec:bulkbulk} we extend this computation into the bulk, building the piecewise mixed-signature spacetime and the bulk-to-bulk propagator for a scalar field. We show how this is related to ingoing boundary conditions on a single copy of a Lorentzian black hole spacetime even for the nonlinear higher-point problem. In section \ref{sec:numerics} we use these results to numerically compute the scalar three-point function in momentum space. In section \ref{sec:TOO} we investigate heating of the system due to scalar driving. We conclude in section \ref{sec:conclusion}.
\\\\
\textbf{Note added:} While finalising this preprint, the preprint \cite{Loganayagam:2022zmq} appeared which has some overlap with our results.

\section{Schwinger-Keldysh and a generating function for retarded correlation functions}
\label{sec:fieldtheory}
In this section we review the Schwinger-Keldysh formalism for computing real-time correlation functions in non-equilibrium field theories at finite temperature, $T=1/\beta$. The review will be brief and focused on the pieces we need to compute expectation values of the R-product \eqref{Rproduct}. We mainly follow \cite{Wang:1998wg} (though note our sign conventions differ in places; we have chosen signs to match conventions in \cite{vanRees:2009rw}), see also \cite{Liu:2018kfw}.

\begin{figure}[h!]
\centering
\includegraphics[width=0.5\columnwidth]{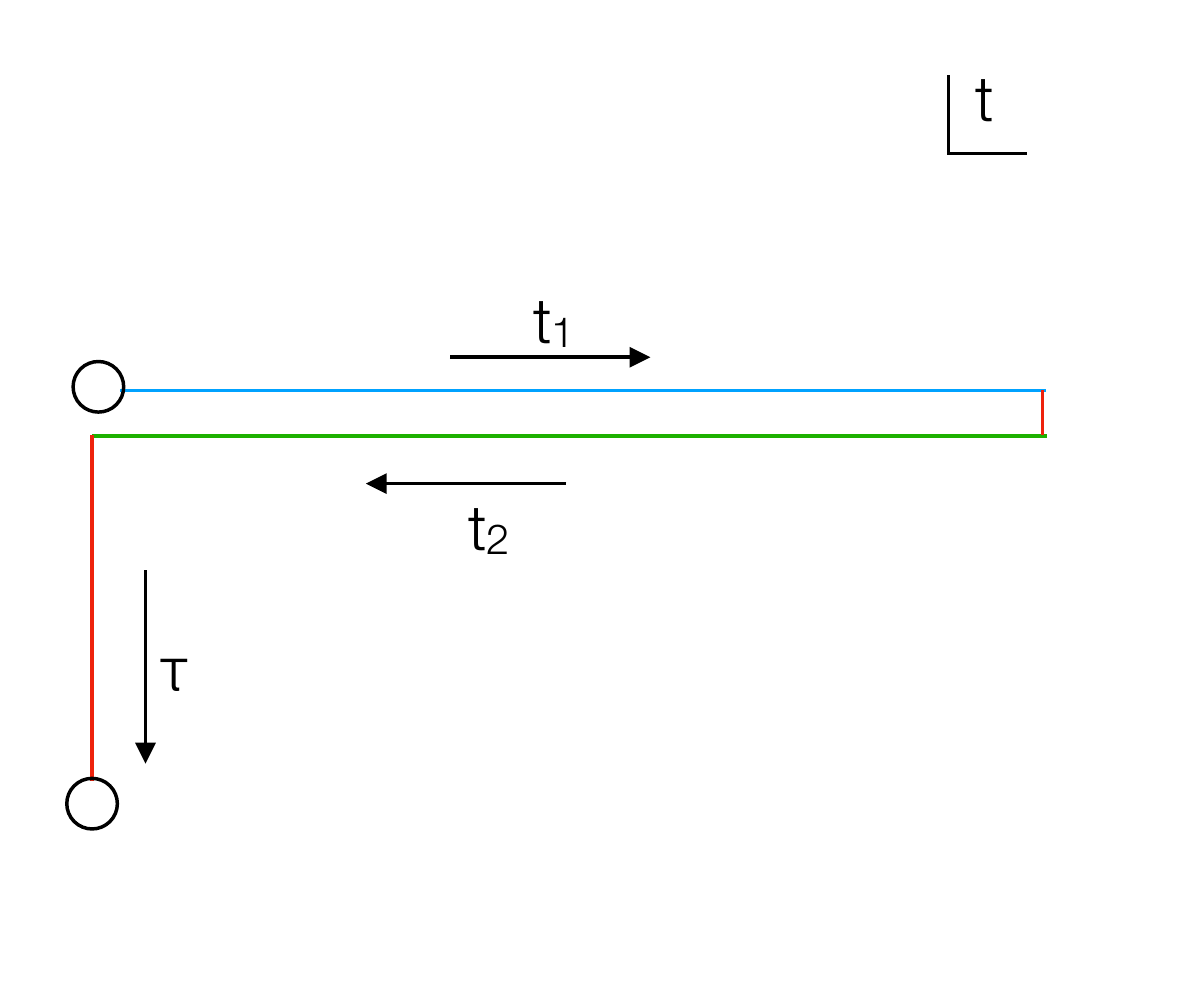}
\put(-150,113){$\mathcal{C}_1$}
\put(-105,87){$\mathcal{C}_2$}
\put(-194,80){$\mathcal{C}_E$}
\caption{The Schwinger-Keldysh contour in the complex time plane. It comprises of two Lorentzian segments labelled $\mathcal{C}_1$ and $\mathcal{C}_2$ on which we have sources $J_1(x)$ and $J_2(x)$, respectively, and a Euclidean part ${\cal C}_E$ on which there is no source. 
The circles are identified to give a closed path. The associated Euclidean periodicity is the inverse temperature $\beta=1/T$. For each segment $t$ is parameterised as follows; on ${\cal C}_1$ we have $t=t_1$ with $t_1\in [0,t_f]$ for some sufficiently large $t_f>0$, on ${\cal C}_2$ we have $t=-t_2$ with $t_2\in [-t_f,0]$, and on ${\cal C}_E$ we have $t = -i \tau$ with $\tau \in [0, \beta]$.}
\label{fig:SK_contour}
\end{figure}

Real-time correlation functions are obtained by evaluating a path integral along the (closed) contour in the complex time plane shown in figure \ref{fig:SK_contour} with sources $J_1(x)$ and $J_2(x)$ (for an operator $\mathcal{O}(x)$) on the two Lorentzian segments of the contour and zero source on the Euclidean part. The operators are also labelled by the segement of the contour on which they are inserted. A convenient basis for the correlation functions is the so-called r/a basis, where the variables are organised in the following way,
\bea
J_a(x)=J_1(x)-J_2(x),\qquad J_r(x)=\frac{1}{2}\left(J_1(x)+J_2(x)\right)\,,\nonumber\\
\mathcal{O}_a(x)=\mathcal{O}_1(x)-\mathcal{O}_2(x),\qquad  \mathcal{O}_r(x)=\frac{1}{2}\left(\mathcal{O}_1(x)+\mathcal{O}_2(x)\right)\,.
\eea
The $n$-point Green's functions for the operator $\mathcal{O}(x)$ are then defined by
\be
G_{\alpha_1\dots\alpha_n}(x_1,x_2\dots,x_{n}) \equiv (-i)^{n-1} 2^{n_r-1}\langle T_p [\mathcal{O}_{\alpha_1}(x_1)\dots\mathcal{O}_{\alpha_n} (x_n)]\rangle,
\ee
where $T_p$ denotes path ordering along the contour, $\alpha_1,\dots,\alpha_n \in \{a,r\}$,  $n_r$ counts the total number of $r$ indices and $\langle\dots\rangle$ denotes the thermal expectation value.
These can be generated by the following path integral on the contour $p$,
\bea
G_{\alpha_1\dots\alpha_n}(x_1,x_2\dots,x_{n}) &=& 2^{n_r-1} i \frac{\delta^n Z}{\delta J_{\bar{\alpha}_1}\ldots \delta J_{\bar{\alpha}_n}}\bigg|_{J_r = J_a = 0},\nonumber\\
Z[J_a(x),J_r(x)]&\equiv& \left< T_p e^{-i \int \left(J_a(y)\mathcal{O}_r(y)+ J_r(y)\mathcal{O}_a(y)\right)}\right>,
\eea
here $\bar{a}=r$, $\bar{r}=a$. Specifically for two-point functions ($n=2$), one has\footnote{In the notation of \cite{vanRees:2009rw} we have $G_{ra}(x_1,x_2) = \Delta_R(x_1,x_2)$.}
\bea
G_{rr}(x_1,x_2)&=&-i\, \langle{\mathcal\{O}(x_1),\mathcal{O}(x_2) \}\rangle\,,\\
G_{ra}(x_1,x_2)&=&-i \,\theta(x_1^0-x_2^0)\langle[\mathcal{O}(x_1), \mathcal{O}(x_2)]\rangle\,,\label{Gra}\\
G_{ar}(x_1,x_2)&=&i \,\theta(x_2^0-x_1^0)\langle[\mathcal{O}(x_1),\mathcal{O}(x_2)]\rangle\,.
\eea
Note that $G_{rr}$,$G_{ra}$,$G_{ar}$ correspond directly to the symmetric, retarded and advanced Green's functions and thus capture the response and fluctuation functions of the system. 
This interpretation of the correlation functions goes beyond two-point functions; the r/a correlation functions capture the full set of time-ordered response and fluctuation functions together with associated generalised fluctuation-dissipation theorems \cite{Wang:1998wg}.\footnote{For generalisations to out-of-time-ordered correlators see for example \cite{Chaudhuri:2018ymp}.} 
For example at the three-point function level ($n=3$) one obtains\footnote{In the notation of \cite{vanRees:2009rw} we have $-G_{raa}(x_1,x_2,x_3) = \Delta(x_1,x_2,x_3) = \Delta_{RR}(x_1,x_2,x_3) + \Delta_{RR}(x_1,x_3,x_2)$.}
\bea
G_{raa}(x_1,x_2,x_3) &=&  -\theta(x_1^0-x_2^0)\theta(x_2^0-x_3^0)\big<[[\mathcal{O}(x_1),\mathcal{O}(x_2)],\mathcal{O}(x_3)]\big> + (2\leftrightarrow 3)\,.\label{Graa}
\eea
For our purposes we note that $G_{ra}, G_{raa}$, and its extension to higher points, $G_{raa\ldots}$ are the expectation values of the R-product we wish to compute \eqref{Rproduct} as shown in \cite{CHOU19851}, i.e. 
\be
G_{ra\ldots a}(x_1,x_2,\ldots, x_n) = \left<R[\mathcal{O}(x_1);\mathcal{O}(x_2)\ldots \mathcal{O}(x_{n})]]\right>.
\ee

For the holographic calculation we carry out later in this paper, instead of using $Z$ to generate these correlation functions, it will be more convenient to use one-point functions in the presence of sources for this role instead. This is because the one-point function is exposed in the bulk geometry as near boundary data (after holographic renormalisation), making the connection more immediate. Furthermore, since we are interested only in the retarded correlators, $G_{ra}, G_{raa}, \ldots$, we can focus on a restricted generating function, namely the one point function with only $J_r$ turned on, $\left<\mathcal{O}_r(x)\right>_J$, obtained from the path integral as follows,
\bea
\left<\mathcal{O}_r(x_1)\right>_J &=&  i \frac{\delta Z[J_a, J_r]}{\delta J_{a}(x_1)}\bigg|_{J_1 = J_2 = J_r}= \left< T \mathcal{O}_r(x_1) e^{-i \int J_r(y)\mathcal{O}_a(y)}\right>\,.
\eea
Here and in what follows, the subscript $J$ indicates that the expectation value is taken in the presence of sources, otherwise it is the expectation value in the thermal state.
Indeed, one can see that this generates the fully retarded correlation functions of interest, since if we expand perturbatively in the forcing $J_r$ we have
\bea
\left< \mathcal{O}_r(x_1)\right>_J &=&  \left< \mathcal{O}_r(x_1)\right> + \int d^dx_2 G_{ra}(x_1,x_2)J_r(x_2) \nonumber\\
&&+ \frac{1}{2} \int d^dx_2 d^dx_3  G_{raa}(x_1,x_2, x_3)J_r(x_2)J_r(x_3) + O(J_r)^3\,.\label{Jexpansion}
\eea
It is these relations that we will use to compute $G_{ra}, G_{raa}$ holographically below.

\section{Holography and ingoing boundary conditions}\label{sec:bulkbulk}
\begin{figure}[h!]
\centering
\includegraphics[width=0.6\columnwidth]{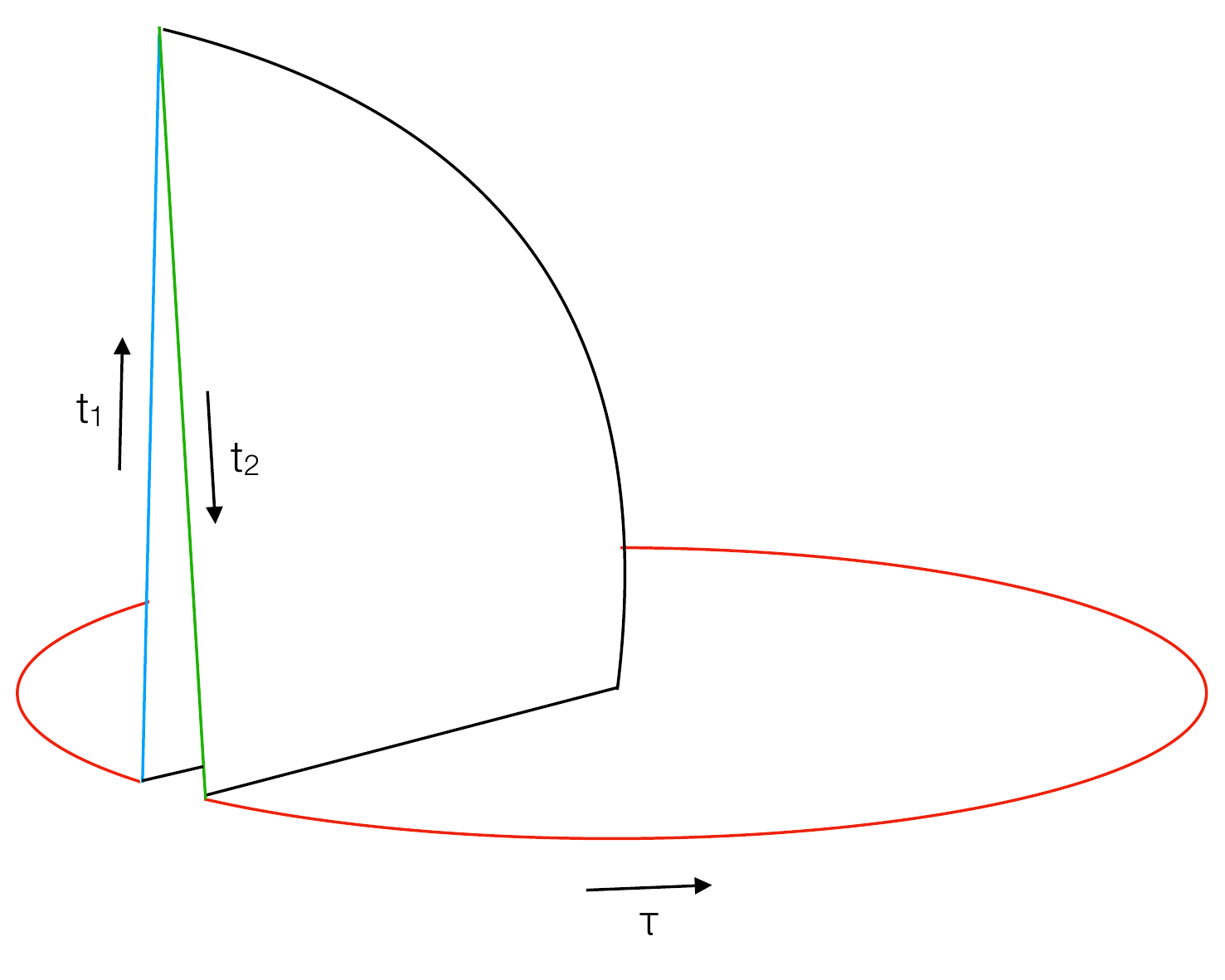}
\put(-238,98){$\mathcal{C}_1$}
\put(-218,128){$\mathcal{C}_2$}
\put(-150,15){$\mathcal{C}_E$}
\put(-175,100){$\mathcal{M}_2$}
\put(-115,50){$\mathcal{M}_E$}
\caption{The bulk geometry $\cal M$ dual to the Schwinger-Keldysh contour in figure \ref{fig:SK_contour}. The Euclidean segment of the contour ${\cal C}_E$ is filled in with a Euclidean black hole; topologically, this fills in the Euclidean time circle to give rise to a (cut) disk, ${\cal M}_E$. The two Lorentzian segments ${\cal C}_1$, ${\cal C}_2$ fill in with two copies of a part of an eternal (Lorentzian) black hole, ${\cal M}_1$, ${\cal M}_2$. The fields are piecewise-smooth, and $C^1$ along the joins between them. }
\label{fig:BulkGeometry}
\end{figure}
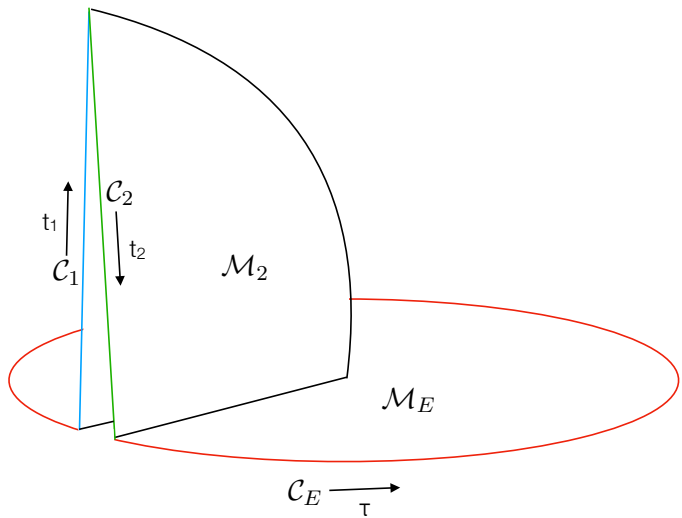

Let us now use the real-time holographic prescription of \cite{Skenderis:2008dg, Skenderis:2008dh} to compute two- and three-point functions for the specific thermal field theory contour shown in figure \ref{fig:SK_contour}. According to this prescription, one needs to fill in the entire field theory contour with bulk spacetimes and solve for the bulk fields subject to the boundary data specified on the contour, and matching conditions on the gluing surfaces of the bulk spacetimes. More specifically, the vertical segment $\mathcal{C}_E$ in figure \ref{fig:SK_contour} is filled in by a Euclidean black hole solution (denoted by $\mathcal{M}_E$) while the two Lorentzian segments, $\mathcal{C}_1$ and $\mathcal{C}_2$, correspond to two copies of the portion of an eternal Lorentzian black hole solution between the $t=0$ slice and some late-time surface $t=t_f$ (denoted by $\mathcal{M}_1$ and $\mathcal{M}_2$ respectively).
The total space, denote by $\mathcal{M}$, is sketched in figure \ref{fig:BulkGeometry}. In terms of the matching conditions, these correspond to continuity of the field and the conjugate momentum at each gluing surface between the various segments of the bulk manifolds (fixed Schwarzschild coordinate $r$ and transverse $x^i$) and can be understood as a $C^1$ gluing of the field.

\subsection{Holographic setup}
The boundary operator $\mathcal{O}$ is dual to an interacting scalar field in the bulk. In particular, consider the bulk action
\bea
S &=& \int d^{d+1}x\sqrt{-g}\left(R  - \frac{1}{2} (\partial\phi)^2 - V(\phi)\right)\\
V &=& 2 \Lambda + \frac{1}{2}m^2 \phi^2 + \lambda \phi^3 + \ldots \quad \text{with} \quad \Lambda = -\frac{d(d-1)}{2L^2}, \quad m^2L^2 = \Delta(\Delta - d)\,.\nonumber
\eea
giving rise to the following equations of motion 
\bea \label{eq:eom}
&R_{\mu\nu}-\frac{1}{2}g_{\mu\nu} V(\phi)-\frac{1}{2}\partial_\mu\phi\partial_\nu\phi =0\,,\nonumber\\
&\Box\phi - \Delta(\Delta -d) \phi=3\,\lambda\phi^2+\dots\,.
\eea
The above equations of motion admit the following solution 
\bea
ds^2&=&-f(r) dt^2+\frac{dr^2}{f(r)}+r^2 dx^i_{d-1} dx^i_{d-1}\,,\qquad f(r)=r^2\left(1-\frac{r_h^d}{r^{d}}\right)\,,\nonumber\\
\phi&=&0\,,\label{eq:Schw}
\eea
which corresponds to the Lorentzian AdS-Schwarszchild black brane with the conformal boundary at $r\to \infty$. The Euclidean version is obtained by continuing $t = -i \tau$.

The thermal state, $J_1=J_2=0$,  we are interested in is dual to the bulk solution \eqref{eq:Schw} and despite being a thermal state, has $\left<\mathcal{O}_r(x_1)\right> = 0$.\footnote{See for example \cite{Myers:2016wsu, Grinberg:2020fdj, Berenstein:2022nlj} for holographic constructions seeking nontrivial thermal one-point functions.} Switching on perturbative sources $J_1, J_2$ on the contour $\mathcal{C}$ as prescribed in figure \ref{fig:BulkGeometry} gives rise to perturbations of the field on the geometry $\mathcal{M}$ which propagate from one segment of the geometry to the others via the matching conditions, which take the form
\bea
\phi_1(t_1=0,x^i,r) &=& \phi_E(\tau=\beta,x^i,r)\label{glue1E}\\
\partial_{t_1}\phi_1(t_1=0,x^i,r) &=& i \partial_\tau\phi_E(\tau=\beta,x^i,r)\label{glued1E}\\
\phi_2(t_2=0,x^i,r) &=& \phi_E(\tau=0,x^i,r)\label{glue2E}\\
\partial_{t_2}\phi_2(t_2=0,x^i,r) &=& -i\partial_\tau\phi_E(\tau=0,x^i,r)\label{glued2E}\\
\phi_1(t_1=t_f,x^i,r) &=& \phi_2(t_2=-t_f,x^i,r)\label{glue12}\\
\partial_{t_1}\phi_1(t_1=t_f,x^i,r) &=& -\partial_{t_2}\phi_2(t_2=-t_f,x^i,r)\,.\label{glued12}
\eea
These can be obtained by demanding $C^1$ continuity at the corners of the contour.

We thus proceed to solve \eqref{eq:eom} perturbatively in the sources on all segments of the bulk manifold. Let $\epsilon$ be a formal parameter which counts the powers of $J_r$. We expand the scalar field and the metric as
\bea
\phi &=& \psi \,\epsilon + \chi\, \epsilon^2 + O(\epsilon)^3,\nonumber\\
g &=& \bar{g} + O(\epsilon)^2.
\eea
Here $\bar{g}$ is the background metric constructed from piecewise-smooth gluing of \eqref{eq:Schw} according to figure \ref{fig:BulkGeometry}. $\psi$ and $\chi$, subject to appropriate boundary conditions described below, will control the two-point function $\left<R[\mathcal{O}(x_1);\mathcal{O}(x_2)]]\right>$ and the three-point function $\left<R[\mathcal{O}(x_1);\mathcal{O}(x_2)\mathcal{O}(x_{3})]]\right>$ respectively. Note that up to this order the metric perturbations decouple and thus we will not need to consider Einstein equations in our perturbative computation; this first appears at order $\epsilon^2$ in $g$. Working up to second order in $\epsilon$ for the scalar, the equation of motion \eqref{eq:eom} gives rise to the following two boundary-value problems. Firstly, we have the boundary-value problem specified by,
\bea
\left(\Box_{\bar{g}_i} - \Delta(\Delta -d)\right)\psi_i &=& 0\,,\nonumber\\
\lim_{r\to\infty} r^{d-\Delta} \epsilon \psi_1&=& J_1,\nonumber\\
\lim_{r\to\infty} r^{d-\Delta} \epsilon \psi_2&=& J_2,\nonumber\\
\lim_{r\to\infty} r^{d-\Delta} \epsilon \psi_E&=& 0,
\label{bvp1}
\eea
subject to regularity in the interior and $C^1$ at the gluing surfaces.
$\Box_{\bar{g}_i}$ denotes the d'Alembertian/Laplacian on the Lorentzian/Euclidean segments of the geometry with metric $\bar{g}_i$ where $i$ labels the segment.
Secondly, given $\psi$, we have our second boundary-value problem
\bea
\left(\Box_{\bar{g}_i} - \Delta(\Delta -d)\right)\chi_i &=&  3\lambda \psi_i^2\,,\nonumber\\
\lim_{r\to\infty} r^{d-\Delta} \chi_i &=& 0 
\label{bvp2}
\eea
subject to regularity in the interior and $C^1$ at the gluing surfaces, where $i=1,2, E$ indicating the three segments of ${\cal M}$.

Given the asymptotic behaviour of $\psi$, $\chi$, we can read off $\left< \mathcal{O}_r(x_1)\right>_J$ using holographic renormalisation, given by $\left<\mathcal{O}_r(x_1)\right>_J = (2\Delta-d)\left(\epsilon\, v^{(\psi)}+\epsilon^2\, v^{(\chi)}\right)$ where $v^{(\psi)}$ and $v^{(\chi)}$ are the coefficients of $r^{-\Delta}$ in the near boundary expansion of $\psi$ and $\chi$ respectively. Then, using \eqref{Jexpansion} we have that,
\be
\left<R[\mathcal{O}(x_1);\mathcal{O}(x_2)\mathcal{O}(x_{3})]]\right> = \frac{\delta^2 \left<\mathcal{O}_r(x_1)\right>_J}{\delta J_r(x_2) \delta J_r(x_3)}\bigg|_{J_r = J_a = 0} = \epsilon^2\,(2\Delta-d)\,\frac{\delta^2\, v^{(\chi)}(x_1)}{\delta J_r(x_2) \delta J_r(x_3)}\bigg|_{J_r = J_a = 0}. \label{R3variations}
\ee
In the following subsection we solve \eqref{bvp1} and \eqref{bvp2} by considering the influence of a single delta function on ${\cal M}_1$ for the entire spacetime ${\cal M}$, and then similarly for a single delta function on ${\cal M}_2$. This gives a matrix of bulk-to-bulk propagators which can be used to generate solutions, focusing on the case $J_1=J_2=J_r$ in order to extract \eqref{R3variations}. We use these solutions to show that one may utilise ingoing boundary conditions on a single copy of the spacetime in order to obtain \eqref{R3variations}. We will then use this result in section \ref{sec:numerics}. 

\subsection{Ingoing conditions from bulk-to-bulk propagators}\label{sec:ingoing}
To solve the boundary value problems \eqref{bvp1} and \eqref{bvp2}, we first construct the bulk-bulk propagator which obeys the matching conditions between $\mathcal{M}_1$, $\mathcal{M}_2$ and $\mathcal{M}_E$. For \eqref{bvp1} we need the bulk-boundary propagator which we can get by a limit of the bulk-bulk.

First consider a delta-function source placed in the bulk on $\mathcal{M}_1$ at $t_1=t_1'$, $x^i={x^i}'$ and at some $r=r'$. The resulting propagator on $\mathcal{M}_1$ is given by
\bea
&\Box_{\bar{g}_1} \Delta_{11}(t_1,x,r;t_1',x',r') = \delta(t_1-t_1')\delta^{(d-1)}(x-x')\delta(r-r')\,,\nonumber\\
&\Delta_{11}(t_1,x,r;t_1',x',r') = \int \frac{d\omega}{2\pi} \frac{d^{d-1}k}{(2\pi)^{d-1}}\, e^{-i \omega t_1+i k x}\left(c_{11}^R G^R(\omega,k,r;r')+c_{11}^A G^A(\omega,k,r;r')\right)\,.
\eea
Here $G^R$,$G^A$ are the bulk scalar retarded and advanced propagators in momentum space, and are normalised so that as $r\to r'$ they satisfy 
\be
G^{R,A}(\omega,k,r;r') = \frac{1}{2}\delta(r-r')\,.
\ee
Thus for a unit strength delta function in $t_1,x$ we have
\be\label{eq:constrM1}
e^{-i \omega t_1'} (c_{11}^R + c_{11}^A) = 1.
\ee
To simplify the notation, in what follows we will suppress the $x$ dependence and the corresponding $k$-integrals \textemdash these are spectators in our calculation and can be reinstated at any stage.

Given the matching condition on the gluing surfaces this field propagates to the other segments of the manifold $\mathcal{M}$. There the propagator has the same functional form since the spacetimes $\bar{g}_i$ are just related under analytically continuing $t$. In particular, on $\mathcal{M}_2$ and $\mathcal{M}_E$ we have
\bea
\Delta_{21}(t_2,r;t_1',r') &=& \int \frac{d\omega}{2\pi} \, e^{i \omega t_2}\left(c_{21}^R G^R(\omega,r;r')+c_{21}^A G^A(\omega,r;r')\right),\\
\Delta_{E1}(\tau,r;t_1',r') &=& \int \frac{d\omega}{2\pi} \, e^{-\omega \tau}\left(c_{E1}^R G^R(\omega,r;r')+c_{E1}^A G^A(\omega,r;r')\right).
\eea
Absence of sources on these segments leads to
\be
c_{21}^R + c_{21}^A = 0,\quad c_{E1}^R + c_{E1}^A = 0.
\ee

We now turn our attention to the matching conditions between various segments of the manifold $\mathcal{M}$ taking advantage of the analytic structure of $G^R$, $G^A$.  For \eqref{glue1E} we have
\be
\int \frac{d\omega}{2\pi} \, \left((e^{i \omega t_1'} -c_{11}^A)G^R+c_{11}^A G^A\right) = \int \frac{d\omega}{2\pi} 
\, e^{-\omega \beta}c_{E1}^A\left(-G^R + G^A\right)\,.\label{eq:matching1E}
\ee
where we have used \eqref{eq:constrM1}. Given that the source has support in the future of the gluing surface and $G^R$ is analytic in the upper-half plane, the term proportional to $e^{i \omega t_1'}$ evaluates to zero. We thus have
\be
\int \frac{d\omega}{2\pi} \,c_{11}^A\left (-G^R+G^A\right) = \int \frac{d\omega}{2\pi} \, e^{-\omega \beta }c_{E1}^A\left(-G^R + G^A\right).\nonumber
\ee
giving
\be
c_{11}^A = e^{-\omega\beta} c_{E1}^A.
\ee
From \eqref{glue2E} we get
\be
c_{21}^A = c_{E1}^A.
\ee
Finally for \eqref{glue12} we have
\be
\int \frac{d\omega}{2\pi}\, e^{-i \omega t_f}\left(c_{11}^R G^R+(e^{i \omega t_1'} -c_{11}^R) G^A\right) = \int \frac{d\omega}{2\pi} \, e^{-i \omega t_f}c_{21}^R\left( G^R- G^A\right).
\ee
In this case the source has support in the past of the gluing surface and $G^A$ is analytic in the lower-half plane allowing us to drop the  contribution proportional to $e^{i \omega t_1'}$ to get
\be
c_{11}^R = c_{21}^R.
\ee
Note that the matching conditions \eqref{glued1E},\eqref{glued2E},\eqref{glued12} involving derivatives do not provide additional constraints on the coefficients. We can summarise the above solution as follows. Let $n =(e^{\beta\omega} -1)^{-1}$, the Bose-Einstein distribution function. Then, taking $t_1'=0$, $x_1'=0$ for presentational simplicity\footnote{Restoring $t_1',x_1'$ follows by rescaling each expression in the following expressions by $e^{i\omega t_1'-i k\cdot x_1'}$}, we have, 
\be
c_{11}^R = 1+n,\, c_{11}^A = -n,\, c_{21}^R= 1+n,\, c_{21}^A = -(1+n),\, c_{E1}^R = 1+n,\,c_{E1}^A=-(1+n).
\ee

Next we construct the solution resulting from a delta function on $\mathcal{M}_2$, which we write as $\Delta_{12}, \Delta_{22}, \Delta_{E2}$ for $\mathcal{M}_1,\mathcal{M}_2,\mathcal{M}_E$ respectively. This gives, by an analogous calculation, 
\be
c_{12}^R = -n,\, c_{12}^A = n,\, c_{22}^R= -n,\, c_{22}^A = 1+n,\, c_{E2}^R = -(1+n),\,c_{E2}^A= 1+n.
\ee
Given the above, we can now construct a solution which has equal sources on $\mathcal{M}_1$ and $\mathcal{M}_2$ by superimposing the above two solutions, which we write as $\Delta_{1r}, \Delta_{2r}, \Delta_{Er}$ for $\mathcal{M}_1,\mathcal{M}_2,\mathcal{M}_E$ respectively. It is given by,
\be
c_{1r}^R = 1,\, c_{1r}^A = 0,\, c_{2r}^R= 1,\, c_{2r}^A = 0,\, c_{Er}^R = 0,\,c_{Er}^A= 0,
\ee
 and is thus vanishing on $\mathcal{M}_E$ and equal on $\mathcal{M}_1$ and $\mathcal{M}_2$ where it is built only from $G^R$.

Let us specialise the above for $\psi$ in \eqref{bvp1}.
By taking an appropriate limit of $r'$ to the boundary this gives us a solution for a delta function source for $J_r$ and thus a solution for $\psi$ in \eqref{bvp1}. Specifically, given the bulk to boundary propagator $\tilde \Delta(t,r;t')$, one has\footnote{In obtaining the second line we have used that $\int_{\partial_r\mathcal{M}_2}dt_2'f(t_2')=\int_{-t_f}^{0} dt'_2\, f(t_2')=\int_0^{t_f} dt'_1\, f(-t_1')=\int_{\partial_r\mathcal{M}_1}dt_1'\, f(-t_1')$. \label{footlabel}}
\bea
\psi_1(t,r) &=& \int dt_1'  \tilde \Delta_{11}(t,r;t_1') J_1(t_1')+\int dt_2'  \tilde \Delta_{12}(t,r;t_2') J_2(t_2') \,,\nonumber\\
&=&\int dt_1' \left(\tilde \Delta_{11}(t,r;t_1')+\tilde \Delta_{12}(t,r;-t_1') \right)\,J_r(t_1')\,,\nonumber\\
&=&\int dt_1' \,\tilde \Delta_{1r}(t,r;t_1')\,J_r(t_1') \,,
\eea
and similarly for $\psi_2(t,x,r)$ and $\psi_E(t,x,r)=0$. Note that $\psi_1(t,x,r)$ receives no contribution from the Euclidean segment of the spacetime given that the source vanishes there. Note also $\psi_1(t,x,r)$ is built entirely from $G^R$, thus when solving for $\psi_1$ we can simply use ingoing boundary conditions at the horizon on a single-sheeted spacetime. Thus, as shown in \cite{vanRees:2009rw}, the real-time prescription reproduces the well-known and successful recipe of \cite{Son:2002sd,Herzog:2002pc}.

Let us specialise the above for $\chi$ in \eqref{bvp2}. in this case we obtain $\chi$ by integrating the bulk to bulk propagator against $3\lambda\psi^2$. In particular, 
\bea
(3\lambda)^{-1}\chi_1(t,r) &=& \int dt_1' dr' \Delta_{11}(t,r;t_1',r') \psi_1^2(t_1',r')+\int dt_2' dr'\Delta_{12}(t,r;t_2',r') \psi_2^2(t_2',r')\,,\nonumber\\
&=& \int dt_1' dr' \Delta_{11}(t,r;t_1',r') \psi_1^2(t_1',r')+\int dt_2' dr'\Delta_{12}(t,r;t_2',r') \psi_1^2(-t_2',r')\,,\nonumber\\
&=& \int dt_1' dr' \left(\Delta_{11}(t,r;t_1',r') +\Delta_{12}(t,r;-t_1',r') \right)\psi_1^2(t_1',r')\,,\nonumber\\
&=& \int dt_1' dr' \Delta_{1r}(t,r;t_1',r')\psi_1^2(t_1',r')\,,
\eea
where in the third line we have used again that $t_2'=-t_1'$ and that the integration over $t_2'$ goes from $-t_f$ to $0$ (see footnote \ref{footlabel}).
In the above we have used that $\psi$ satisfies $\psi_1(t_1) = \psi_2(-t_1)$, and hence the bulk sources for $\chi$ on the right hand side of \eqref{bvp2} are equal on $\mathcal{M}_1$ and $\mathcal{M}_2$ (and vanish  on $\mathcal{M}_E$). Hence, $\chi$ can be built from $\Delta_{1r}, \Delta_{2r}, \Delta_{Er}$, satisfying $\chi_E=0$, $\chi_1(t_1) = \chi_2(-t_1)$ and built entirely from $G^R$. A subtlety here is that we have non-zero source $\psi^2$ on the gluing surface itself between $\mathcal{M}_1$ and $\mathcal{M}_2$, which violates one of the key steps we took in obtaining the above solution when solving \eqref{glue12}. However, it is easy to see that since $c_{1r}^R = c_{2r}^R$ and $c_{1r}^A = c_{2r}^A$, \eqref{glue12} is actually satisfied in this particular case and thus our analysis still holds.

The remaining question is whether we can obtain $\chi$ with a simple boundary condition in momentum space on a single Lorentzian segment of the spacetime. Since $\chi$ is obtained from only $G^R$ in the bulk, it follows that the response is causal. In linear response this would be sufficient for concluding that the field is ingoing at the black hole horizon on ${\cal M}_1$ (reached as $t_f\to\infty$). We present here an argument in support of this holding at all orders in the $\epsilon$ expansion. We start by considering the near-horizon behaviour of a single Fourier mode (say $k=0$) of the field $\phi$ in Schwarzschild coordinates\footnote{For a probe field $\alpha=1/(4 \pi T)$, but with perturbative backreaction $\alpha$ may receive corrections.}
\be
\phi_\omega = e^{-i \omega t } (r-r_h)^{\pm i \alpha\omega}\left(a_H(\omega)+\ldots\right), \quad \text{for $r>r_h$}
\ee
 Here  the $\pm$ correspond to outgoing and ingoing modes respectively. The form of this expansion holds for both $\psi$ and for $\chi$. In position space,
\bea
\phi(t,r) = \int \phi_\omega d\omega &=& \int e^{-i \omega t } (r-r_h)^{\pm i \alpha \omega}a_H(\omega) d\omega + \ldots\\
&=&\int e^{-i \omega ( t \mp  \alpha \log(r-r_h))}a_H(\omega) d\omega + \ldots.
\eea
As we approach the horizon from the outside, $r\to r_h^+$, the exponential term oscillates rapidly in $\omega$ except near $t = \pm \alpha \log(r-r_h)$. Thus, provided that $a_H(\omega)$ is bounded for real $\omega$, this determines the support of the function in $t$. In particular, support is in the past ($t<0$) for the outgoing case and in the future ($t>0$) for the ingoing case. Thus if we specify anything other than purely ingoing boundary conditions, the solution will have support in the past, and thus not be constructed purely from $G^R$.

\section{Explicit three-point function computation in AdS$_5$/CFT$_4$}
\label{sec:numerics}
With the result of section \ref{sec:bulkbulk} we have established that the PDE problems \eqref{bvp1} and subsequently \eqref{bvp2} can be solved by restricting to a single section of Lorentzian spacetime and imposing ingoing boundary conditions. In this section we utilise this result to compute three-point functions numerically in momentum space for the case of a CFT$_4$ with an asymptotically AdS$_5$ dual.

We will compute the momentum space expectation value of the R-product, given by\footnote{Without loss of generality we have placed one point at $x=0$ using translation invariance. Relaxing this restores the momentum conservation delta function,
\bea
\int d^dx \,d^dx_1 \dots d^d x_{n-1} \left<R[\mathcal{O}(x);\mathcal{O}(x_1)\dots \mathcal{O}(x_{n-1})]\right> e^{i( p\cdot x +p_1\cdot x_1 +\dots + p_{n-1}\cdot x_{n-1})} = \qquad \qquad \qquad &\nonumber\\
(2\pi)^d \delta^{(d)}(p+p_1+\dots+p_{n-1})\,r_n(p_1,\dots,p_{n-1}). &\nonumber
\eea
}
\be
r_n(p_1,\dots,p_{n-1}) \equiv \int d^dx_1 \dots d^d x_{n-1} \left<R[\mathcal{O}(0);\mathcal{O}(x_1)\dots \mathcal{O}(x_{n-1})]\right> e^{i (p_1\cdot x_1 +\dots + p_{n-1}\cdot x_{n-1})}.\label{Rmomentum}
\ee
The R-product appearing in \eqref{Rproduct} is only non-vanishing if the $x_i$ lie inside the past lightcone of $x=0$. This guarantees certain analyticity properties of $r_n$. In particular $r_n(p_1,\dots,p_{n-1})$ is analytic in $p_i$ for $\text{Im}(p_i)$ inside the future lightcone; see \cite{StreaterAndWightman, Haag:1992hx} and the discussion in \cite{Meltzer:2021bmb}. 
In the case of interest we will consider real spatial momenta, and hence $\text{Im}(p_i) = (\text{Im}(p^0_i), 0,\ldots, 0)$. Hence analyticity is guaranteed where all $p^0_i$ are simultaneously in the upper-half frequency plane. 

In the case of three points, $n=3$, and generic external driving momenta $p_i$, the three-point function describes a forced excitation of the scalar field with bulk tree-level diagrams mediated by $\lambda \phi^3$. Let us discuss the expected analytic structure of the corresponding correlator $r_3(p_1,p_2)$. We expect various singularities when the external legs correspond to QNMs. Denote the QNM momenta as $\mathfrak{p}_i$. Such singularities should therefore occur at,
\bea
p_1 = \mathfrak{p}_i \quad \text{or}\quad p_2 = \mathfrak{p}_i \quad \text{or} \quad p_1 + p_2 = \mathfrak{p}_i,\label{threepointsingle}
\eea
for any QNM label $i$. The first two conditions in \eqref{threepointsingle} correspond to external legs of the diagram being driven at QNM frequencies, while the last condition in \eqref{threepointsingle} corresponds to a resonant excitation of a QNM where the total incoming momentum is a QNM and thus the outgoing leg has QNM momentum. If only one of these conditions hold then we anticipate simple poles. Poles of order two can always be arranged kinematically by choices of the $p_i$ such that two of these conditions hold simultaneously. Order three poles are expected if, in the theory under consideration, all three conditions can hold. In more detail, let $p_1 = (\mathfrak{w}_i(|\vec{k}_1|), \vec{k}_1)$, $p_2 = (\mathfrak{w}_j(|\vec{k}_2|), \vec{k}_2)$ where $\mathfrak{w}_i$ denotes a QNM dispersion relation and $\vec{k}_1, \vec{k}_2 \in \mathbb{R}^{d-1}$. Then satisfying all three conditions in \eqref{threepointsingle} requires some choice of $i,j,k$ and $\vec{k}_1, \vec{k}_2$ such that
\be
\mathfrak{w}_i\left(|\vec{k}_1|\right) + \mathfrak{w}_j\left(|\vec{k}_2|\right) = \mathfrak{w}_k\left(|\vec{k}_1+\vec{k}_2|\right). \label{qnmoverlap}
\ee
This is not guaranteed to hold, and whether or not it does likely depends on the details of the theory.

At other generic values of momenta not satisfying any of \eqref{threepointsingle}, one can think about the driving momenta $p_i$ exciting a normalisable bulk field which is not related to any QNM. In the literature such excitations are known as forced modes \cite{Ioka:2007ak}.

\subsection{Plane wave perturbations}
Proceeding with the bulk computation, we decompose the fields $\psi, \chi$ using two plane waves with independent, formal amplitudes $s_1, s_2$ as follows,
\bea
\psi &=& s_1 \psi_1(r) e^{i p_1\cdot x} + s_2 \psi_2(r) e^{i p_2\cdot x},\\
\chi &=& s_1^2 \chi_{11}(r) e^{2 i p_1\cdot x} + s_1 s_2 \chi_{12}(r) e^{i (p_1+p_2)\cdot x} + s_2^2 \chi_{22}(r) e^{2 i p_2\cdot x}.
\eea
At the black hole horizon we impose ingoing boundary conditions on all fields, which is equivalent to constructing the solution on the full piecewise mixed-signature spacetime appropriate for the computation of $r_3(p_1,p_2)$, as detailed in section \ref{sec:bulkbulk}. 
Near the boundary we impose that $\psi_i(r)$ are non-normalisable with unit-coefficient, and that $\chi_{ij}(r)$ are normalisable,
\bea
\psi_i &=& \frac{1}{r^{d-\Delta}} + \ldots + \frac{\psi_i^{(\Delta)}}{r^\Delta} + \ldots,\label{nbpsi}\\
\chi_{ij} &=& \frac{\chi_{ij}^{(\Delta)}}{r^\Delta} + \ldots.
\eea
With this near-boundary behaviour, the CFT source and vev in the presence of sources is given by
\begin{align}
J_r(x) &= s_1 e^{i p_1\cdot x} + s_2 e^{i p_2\cdot x},\\
\left<\mathcal{O}_r(x)\right>_J &=  (2\Delta - d)\bigg[ s_1 \psi_1^{(\Delta)} e^{i p_1\cdot x} + s_2 \psi_2^{(\Delta)} e^{i p_2\cdot x}\nonumber\\
&\qquad\qquad\qquad+ s_1^2 \chi_{11}^{(\Delta)} e^{2i p_1\cdot x} +  s_1 s_2 \chi_{12}^{(\Delta)}  e^{i (p_1+p_2)\cdot x} + s_2^2 \chi_{22}^{(\Delta)} e^{2i p_2\cdot x}\bigg].
\end{align}
Plugging these expressions into \eqref{Jexpansion} and matching coefficients of $s_1$ and $s_1 s_2$ we arrive at the desired expressions for the retarded correlation functions,
\bea
r_2(p_1) &=& (2\Delta - d)\psi_1^{(\Delta)},\\
r_3(p_1,p_2) &=& (2\Delta - d)\chi_{12}^{(\Delta)}.\label{r3final}
\eea
Given a choice of $p_1, p_2$ we solve the boundary value problems for $\psi_i(r)$ and $\chi_{12}(r)$ as described above. The solution gives us the vev data $\chi_{12}^{(\Delta)}$ and hence $r_3(p_1,p_2)$ through \eqref{r3final}. 

For completeness we now provide details of the numerical algorithm used to solve for $\psi_i(r), \chi_{12}(r)$.
The required solutions for $\psi_i(r), \chi_{12}(r)$ can be obtained for a given choice of $p_1,p_2$ straightforwardly by a standard shooting method. However, a faster algorithm requiring no root finding is as follows: integrate each $\psi_i(r)$ from the horizon given ingoing boundary conditions there. At the boundary, the non-normalisable coefficient in $\psi_i(r)$ can be read off, $\psi_i^{(d-\Delta)}$. This will not obey the unit boundary condition required \eqref{nbpsi} but we can compensate for it later. 
Similarly, integrate $\chi_{12}(r)$ from the horizon with ingoing boundary conditions, with $\lambda\neq 0$ to obtain a particular solution $\chi_{12\text{p}}(r)$ and again with $\lambda=0$ to obtain a homogeneous solution, $\chi_{12\text{h}}(r)$. With an appropriate choice of $\alpha$ the required normalisable solution is thus obtained, $\chi_{12}(r)=\chi_{12\text{p}}(r) - \alpha\, \chi_{12\text{h}}(r)$. The vev portion of $\chi_{12}(r)$ can then be rescaled appropriately to compensate for the incorrect strength of the non-normalisable coefficient of $\psi_1(r)$, i.e. $\chi_{12}^{(\Delta)}\to \chi_{12}^{(\Delta)}/(\psi_1^{(d-\Delta)}\psi_2^{(d-\Delta)})$, yielding the three-point function via \eqref{r3final}.

\subsection{Numerical results}
\begin{figure}[h!]
\begin{center}
    \includegraphics[width=0.7\columnwidth]{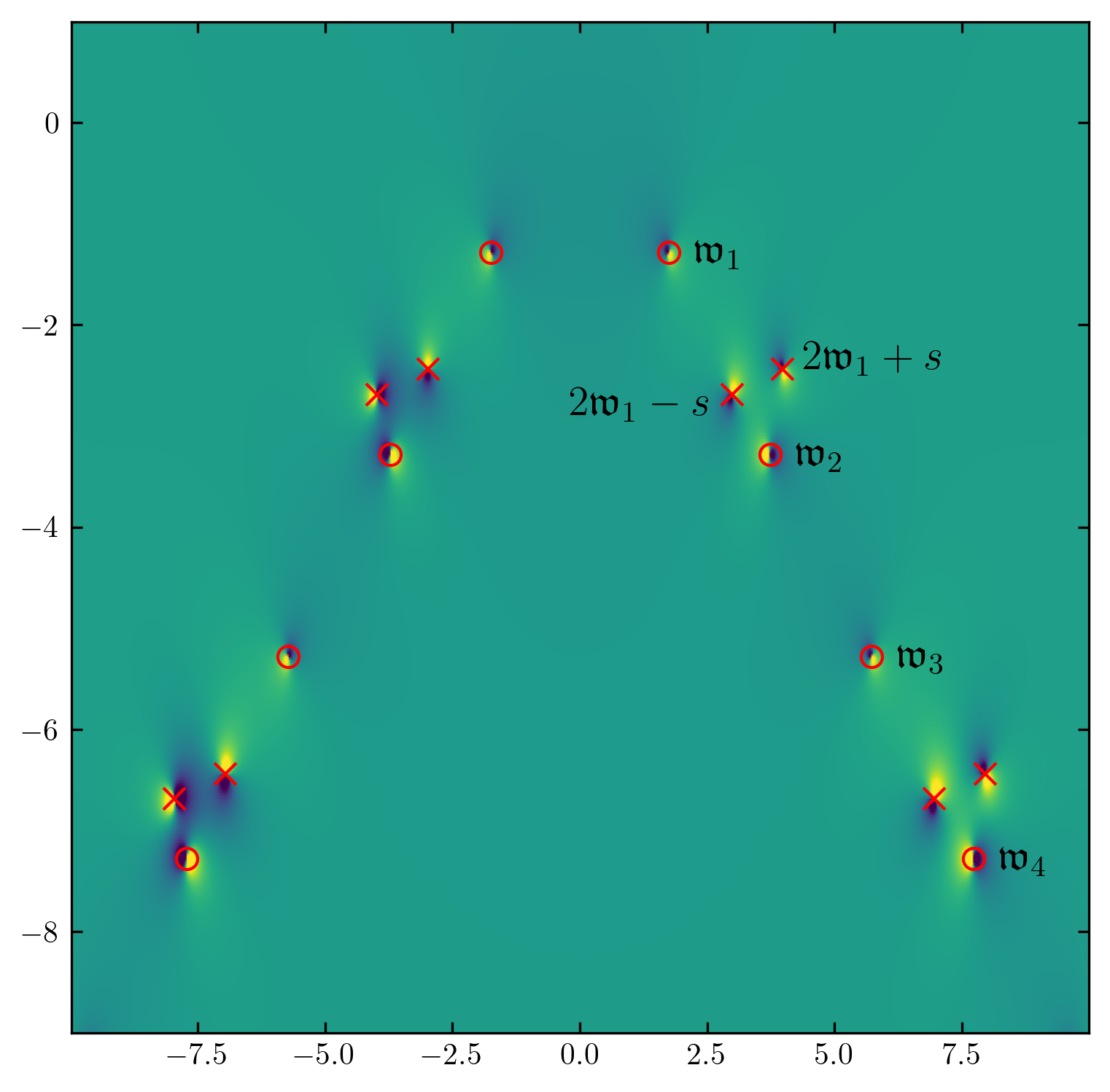}
    \caption{The retarded three-point function, $\text{Re}\,r_3(p_1,p_2)$, shown at zero spatial momentum $p_1^i = p_2^i = 0$ in the complex plane of the total driving frequency, $p = p_1^0+p_2^0$, with an arbitrary fixed frequency difference $s \equiv p_1^0-p_2^0$ (here, for illustration $s=1/2+i/8$). All singularities shown are simple poles, each corresponding to a single leg of the three-point function taking a QNM frequency, denoted by $\mathfrak{w}_i$. Circles correspond to resonant excitations of QNMs through bulk interactions, while crosses correspond to direct driving of QNMs by the external momenta. By tuning $s$ the crosses can be made to collide with other singularities, this is demonstrated in figure \ref{fig:special}.}
    \label{fig:generic}
\end{center}
\end{figure}

\begin{figure}[h!]
\begin{center}
    \includegraphics[width=0.7\columnwidth]{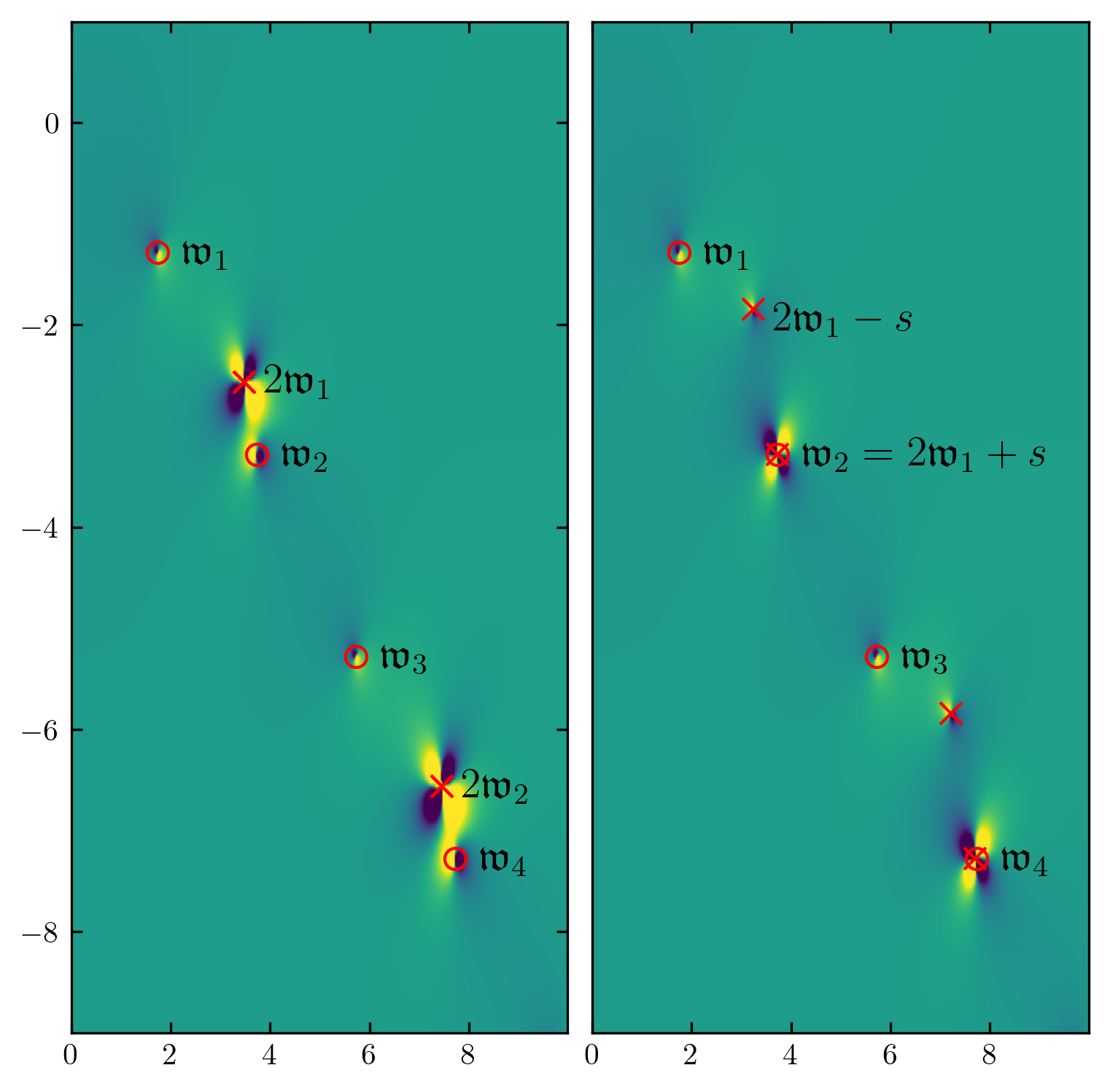}
    \caption{
    The retarded three-point function, $\text{Re}\,r_3(p_1,p_2)$, shown at zero spatial momentum $p_1^i = p_2^i = 0$ in the complex plane of the total driving frequency, $p = p_1^0+p_2^0$, with a carefully chosen fixed frequency difference $s \equiv p_1^0-p_2^0$. Circles correspond to resonant excitations of QNMs through bulk interactions, while crosses correspond to direct driving of QNMs by the external momenta.
    \textbf{Left panel}: $s=0$. Here all QNMs excited by direct driving collide with each other giving a sequence of order-two poles, where the two incoming legs of the diagram are on QNM frequencies.
    \textbf{Right panel}: $s=\mathfrak{w}_2-2\mathfrak{w}_1$ which has been chosen to illustrate another sort of singularity, where a directly driven mode collides with a resonant excitation at $p=\mathfrak{w}_2$ (i.e. at $p_1 = \mathfrak{w}_2 - \mathfrak{w_1}, p_2 = \mathfrak{w}_1$). Thus here the outgoing leg is at $\mathfrak{w}_2$ and one incoming leg is at $\mathfrak{w}_1$. Note that apparent coincidence of poles near $\mathfrak{w}_4$ is only approximate; this is still a pair of simple poles.
    }
    \label{fig:special}
\end{center}
\end{figure}

The results are shown in figures \ref{fig:generic} and \ref{fig:special} for the case of $d=4$, $\Delta = 5/2$, $\lambda = 1$ with all spatial momenta set to zero $\vec{k}_1 = \vec{k}_2 = 0$. In figure \ref{fig:generic} we show the correlator with unequal driving momenta, $r_3((p+s)/2,(p-s)/2)$, in the complex frequency $p$ plane, with fixed complex frequency difference, $s\neq 0$. In this generic situation, by scanning over $p\in \mathbb{C}$ only one of the three possible singularity conditions \eqref{threepointsingle} hold at any one point. The results conform to these expectations; we confirm single poles at these locations consistent with the resonant or direct driving of QNM legs. In particular, a subset of the singularities is the familiar `christmas tree' of simple poles, however now these correspond to resonant excitations of QNMs due to interactions. In figure \ref{fig:special} we fine tune the frequency difference $s$ to guarantee that two of the conditions \eqref{threepointsingle} hold at some $p\in \mathbb{C}$. In the left panel this is $s=0$ giving infinitely many order two poles where both incoming legs are on QNM frequencies, while in the right panel we show a different type of collision where one incoming and one outgoing are on (different) QNM frequencies.

We note that as an additional check, our results are consistent with the analyticity properties of $r_3(p_1,p_2)$ discussed above, following from the causal nature of the correlator. The resonant poles of $r_3(p_1,p_2)$ occur whenever $p_1^0+p_2^0 = \mathfrak{w}_i$, and since $\text{Im}(\mathfrak{w}_i)<0$ it is never the case that this pole is located such that $p_1^0$ and $p_2^0$ are in the upper-half plane at the same time. The driven poles occur at $p_1^0 = \mathfrak{w}_i$ or $p_2^0 = \mathfrak{w}_i$, and again since $\text{Im}(\mathfrak{w}_i)<0$ this pole is located where one or the other of $p_1,p_2$ is in the lower half plane. Note however it is permissible, and is indeed the case, that the driven pole can appear in the upper half plane of $p_1^0 + p_2^0$. To get it there requires driving with an exponentially growing source and hence does not represent a dynamical instability.

\section{$\left<R[T_{\mu\nu}(x);\mathcal{O}(x_1)\mathcal{O}(x_2)]\right>$ from Ward identities}
\label{sec:TOO}
In addition to the scalar operators $O_i$ on each segment of the complex contour ${\cal C}_i$ we also consider the similarly-labelled stress tensor operator, $(T_i)_{\mu\nu}$. The expectation values of these operators in the presence of the scalar sources $J_i$ are related by the CFT trace and diffeomorphism Ward identities which hold locally, 
\bea
\eta^{\mu\nu}\left<(T_i)_{\mu\nu}\right>_J &=& (d-\Delta)J_i\left< O_i \right>_J, \label{Ward:trace}\\
\partial^\mu \left<(T_i)_{\mu\nu}\right>_J &=& \partial_\nu J_i \left< O_i \right>_J.\label{Ward:diff}
\eea
We can use these relations to compute retarded correlation functions with stress tensor insertions in arbitrary CFTs. In particular, we specialise \eqref{Ward:trace} and \eqref{Ward:diff} to the case where $J_1 = J_2$. In the r/a notation this gives $J_r = J_1 = J_2$ and $J_a = 0$. Then, by summing both equations over both Lorentzian segments, we obtain 
\bea
\eta^{\mu\nu}\left<(T_r)_{\mu\nu}\right>_J &=& (d-\Delta)J_r\left<\mathcal{O}_r \right>_J,\\
\partial^\mu \left<(T_r)_{\mu\nu}\right>_J &=& \partial_\nu J_r \left<\mathcal{O}_r\right>_J.
\eea
Next we can expand these one-point functions in the scalar source $J_r$,
\bea
\left<(T_r)_{\mu\nu}\right>_J(x) &=& \left<(T_r)_{\mu\nu}\right>(x) + \frac{1}{2}\int d^dx_1 d^dx_2 (G_{raa}^{TOO})_{\mu\nu}(x,x_1,x_2) J_r(x_1) J_r(x_2) + \ldots,\\
\left<\mathcal{O}_r\right>_J(x) &=& \int d^dy G_{ra}(x,x_1) J_r(x_1) + \ldots\,.
\eea
Where we have adopted the notation $(G_{raa}^{TOO})_{\mu\nu}(x,x_1,x_2) =  \left<R[T_{\mu\nu}(x);\mathcal{O}(x_1)\mathcal{O}(x_2)]\right>$.
Plugging these expressions into \eqref{Ward:trace} and \eqref{Ward:diff} and expanding to quadratic order in $J_r$ we find the following position space relations,
\bea
\eta^{\mu\nu}(G_{raa}^{TOO})_{\mu\nu}(x,x_1,x_2) &=& (d-\Delta) G_{ra}(x,x_1) \delta^{(d)}(x_2-x) + (1\leftrightarrow 2),\\
\frac{\partial}{\partial x_\mu} (G_{raa}^{TOO})_{\mu\nu}(x,x_1,x_2)  &=&  -G_{ra}(x,x_1) \frac{\partial}{\partial x_2^\nu}\delta^{(d)}(x_2-x)+(1\leftrightarrow 2).
\eea
and by \eqref{Rmomentum} we obtain the momentum space results,
\bea
\eta^{\mu\nu}(r^{TOO}_3)_{\mu\nu}(p_1,p_2) &=& (d-\Delta) (r_2(p_1)  + r_2(p_2)),\\
(p_1+p_2)^\mu (r^{TOO}_3)_{\mu\nu}(p_1,p_2)  &=&  r_2(p_1) (p_2)_\nu + r_2(p_2) (p_1)_\nu. \label{wardmom2}
\eea
\sloppy Thus the trace and $(p_1+p_2)$-longitudinal part of the retarded three-point function $\left<R[T_{\mu\nu}(x);\mathcal{O}(x_1)\mathcal{O}(x_2)]\right>$ is determined by the retarded scalar two-point function $\left<R[\mathcal{O}(x);\mathcal{O}(x_1)]\right>$. 
This procedure can be straightforwardly extended to higher scalar points where the $n$-point function $\left<R[T_{\mu\nu}(x);\mathcal{O}(x_1)\ldots \mathcal{O}(x_{n-1})]\right>$ is similarly determined by the $(n-1)$-point function $\left<R[\mathcal{O}(x);\mathcal{O}(x_1)\ldots \mathcal{O}(x_{n-2})]\right>$.
In the the case where the spatial momenta are zero, $p_i = (\omega_i,0)$, \eqref{wardmom2} can be solved explicitly to obtain,
\be
(r^{TOO}_3)_{tt}(\omega_1, \omega_2)  = - \frac{r_2(\omega_1) \omega_2 + r_2(\omega_2) \omega_1}{\omega_1+\omega_2}. \label{Joule}
\ee
Because we only used the Ward identities to obtain these results, they hold generally for CFTs.

The quantity \eqref{Joule} encodes how the system heats up under external driving at frequencies $\omega_i$, and thus is a scalar analogue of Joule heating. The Joule heating effect is similarly determined by Ward identities where instead of $\partial_\mu \left<O\right>_J$ on the right hand side of \eqref{Ward:diff}, one has terms associated to the $U(1)$ current, $F_{\nu\mu}\left<J^\mu\right>_J$, and the associated three-point function $\left<R[T_{tt}(x);J_x(x_1)J_x(x_2)]\right>$ is determined in terms of $\left<R[J_x(x);J_x(x_1)]\right>$ and thus the conductivity of the system. Joule heating in holography has been explored in \cite{Horowitz:2013mia, Withers:2016lft}. 

\section{Discussion}
\label{sec:conclusion}

In this work we computed retarded three-point functions of holographic QFTs by solving for interacting scalar fields propagating on the geometry dual to the SK contour. We showed how this construction is analogous to requiring ingoing boundary conditions on a Lorentzian black hole spacetime. We analysed the analytic structure and found singularities corresponding to the interactions of QNMs in the bulk mediated by three-point couplings.

At two points there are three non-trivial correlators on the SK contour; the retarded $G_{ra}$ (the expectation value of the R-product), the advanced $G_{ar}$ and fluctuations $G_{rr}$. $G_{aa}$ vanishes identically. However these are not all independent. In momentum space, $\tilde{G}_{ra}$ and $\tilde{G}_{ar}$ are related through complex conjugation while $\tilde{G}_{rr}$ is given by the fluctuation-dissipation theorem,
\be
\tilde{G}_{rr} = (1+2n)\left(\tilde{G}_{ra} - \tilde{G}_{ar}\right),
\ee
where $n$ is the Bose-Einstein distribution function introduced earlier. Holographically, this identity is manifest in the bulk-bulk propagators we derived in section \ref{sec:ingoing} since one may verify that
\be
\Delta_{rr} = (1+2n)\left(\Delta_{ra}-\Delta_{ar}\right),
\ee
where $\Delta_{rr} = \frac{1}{2}\left(\Delta_{11}-\Delta_{12}+\Delta_{21}-\Delta_{22}\right)$, $\Delta_{ra} = \frac{1}{2}\left(\Delta_{11}+\Delta_{12}+\Delta_{21}+\Delta_{22}\right)$, $\Delta_{ar} = \frac{1}{2}\left(\Delta_{11}-\Delta_{12}-\Delta_{21}+\Delta_{22}\right)$ which follow from the change of basis. For completeness we note that $\Delta_{aa} = \frac{1}{2}\left(\Delta_{11} + \Delta_{12} - \Delta_{21} - \Delta_{22}\right) = 0$ as required.
So at $n=2$ points there is only one independent component of the matrix of correlation functions $G_{
\alpha_1\alpha_2}$. For $n$-points there are $2^{n-1}-1$ independent components of $G_{\alpha_1\ldots\alpha_n}$ owing to higher-order fluctuation dissipation relations \cite{Wang:1998wg}.
At three points in this work we have focused on $G_{raa}$ (the expectation value of the R-product), but it would be interesting to also compute the remaining two independent components, associated fluctuation-dissipation relations and physical interpretation from the bulk perspective.

We also computed three-point correlators involving single stress-tensor insertions. More generally our work serves as a precursor for computing retarded three-point functions of conserved currents. This may be of relevance in experimental domains where nonlinear response properties of currents are under consideration, for example \cite{2020PhRvX..10a1053L}.

\begin{acknowledgments}
It is a pleasure to thank Felix Haehl, Zezhuang Hao, Balt van Rees and Vaios Ziogas for discussions. We would like to acknowledge the Nordita scientific program ``Recent developments in strongly correlated quantum matter'' where this work was initiated. C.P. acknowledges support from a Royal Society - Science Foundation Ireland University Research Fellowship via grant URF/R1/211027. B.W. is supported by a Royal Society University Research Fellowship and in part by the Science and Technology Facilities Council (Consolidated Grant “Exploring the Limits of the Standard Model and Beyond”). 
\end{acknowledgments}

\bibliographystyle{ytphys}
\bibliography{refs}

\end{document}